# Nanoscale heterogeneous phase separation kinetics in binary mixtures: Multistage dynamics

Milan K. Hazra, Sarmistha Sarkar and Biman Bagchi[*]

*Solid State and Structural Chemistry Unit, Indian Institute of Science, Bangalore 560012, India.*
[*]E-mail: bbagchi@sscu.iisc.ernet.in; profbiman@gmail.com

## Abstract

*In order to find a measure of the dynamical features of phase separation kinetics during spinodal decomposition of a liquid binary mixture (like water and cyclohexane, water and 2,6 lutidiene or methanol and cyclohexane), we study both the initial fast exponential-like growth (the Cahn-Hilliard regime) and the subsequent cross-over to a much slower, non-exponential long time growth (the so-called scaling regime), by atomistic molecular dynamics (MD) simulation of a structure breaking binary liquid mixture. In particular, we combine our MD simulations with a coarse grained multi scale modelling (CGMSM) capable of capturing both length and time scales of phase separation kinetics within simulation box. The system is quenched from a higher temperature ($T^* = 5$) to two lower temperatures ($T^* = 0.5$ and $T^* = 0.9$) well below the coexistence temperature $T^* = (1.6 \pm 0.2)$ of the phase diagram. We observe a multiscale phase separation dynamics. Initially the growth is exponential up to a regime of 80-100ps having strong dependence over quench depth. Subsequently a cross-over regime appears where the dynamics slows down considerably due to coarsening through a power law phase. For deeper quench power law growth dominates over the initial exponential and the cross-over becomes transient. We find that for the present parameter values for the binary mixture, the initial rapid growth of structure formation is practically over within 200 ps which is followed by slow structural coarsening. When scaled by the respective viscosities, this time translates to 50-200 ns for water-lutidine binary mixture. The last part of dynamics may extend into ms. CGMSM studies reveal that the phase separation kinetics is heterogeneous; the kinetics, even the nature, of initial phase separation are found to exhibit different microscopic timescales in different regions of the system depending on the initial composition of the probed region. The dynamics of phase separation is slowest in regions that have*

*equitable distribution of the two species, and one can observe the signatures of "up-hill diffusion" that is a trade mark of spinodal decomposition. Phase separation dynamics is found to slow down considerably when the final quenched temperature is moved closer to the estimated consolute point. The simulation results are found to be in good agreement with the theoretical analysis of Kawasaki and Suzuki in the intermediate-to-long time regime of phase separation kinetics.*

## I.    INTRODUCTION

Spinodal decomposition is defined as a time dependent process where an initially homogeneous binary mixture, subsequent to a deep temperature quench, separates into two pure phases **[1-6]**. This is a non-equilibrium process that proceeds through formation of strip-like regions, seen in many places. Because of the importance of spinodal decomposition both as a natural phenomena and also occurrence in metallurgical processes, the process has drawn considerable attentions over the years in solid state chemistry, metallurgy, geology and in several other fields.

Although the systems involved in spinodal decomposition can be chemically diverse and complex, the initial state at high temperature is usually a homogeneous mixture of two simple chemical species denoted here by A (solvent) and B (solute) and the final state in low temperature is separated into two well-defined (A-rich and B-rich) phases. According to the well-known Cahn-Hilliard theory, a single order parameter $\phi(r,t)$ (defined below) can describe the ordering process. This theory considers the stability of the homogeneous system to infinitesimal composition fluctuation, quantified by the position $(r)$ and time (t) dependent order parameter $\phi(r,t)$ represented by,

$$\phi(r,t) = x_A(r,t) - x_B(r,t) \tag{1}$$

where $x_A(\mathbf{r},t)$ and $x_B(\mathbf{r},t)$ are the mole fractions for solvent and solute, respectively.

The time dependence of $\phi(\mathbf{r},t)$ can be expressed through a continuity equation given as,

$$\frac{\partial \phi(\mathbf{r},t)}{\partial t} = -\nabla \cdot \mathbf{J}(\mathbf{r},t) \tag{2a}$$

where $\mathbf{J}$ is the flux of the system. The flux $\mathbf{J}$ in turn is given by,

$$\mathbf{J} = -\frac{D}{k_B T} \nabla \frac{\delta F[\phi(\mathbf{r})]}{\delta \phi(\mathbf{r})} \tag{2b}$$

where D is the diffusion coefficient.

The coarse grained free energy functional in Ginzberg-Landau form can be written as,

$$F[\phi(\mathbf{r})] = \int d\mathbf{r} \left[ \frac{1}{2} K (\nabla \phi)^2 + f(\phi) \right] \tag{2c}$$

Here $f(\phi)$ denotes the Landau free energy functional that exhibits one minimum at high temperature but two minima separated by a barrier at low temperature.

We combine Eqs. (1) and (2) to obtain the equation of motion for the order parameter as,

$$\frac{\partial \phi(\mathbf{r})}{\partial t} = D \nabla^2 \left[ -K \nabla^2 \phi(\mathbf{r}) + \frac{\partial f}{\partial \phi} \right]. \tag{3}$$

This is essentially generalized diffusion equation, sometimes referred to as Cahn-Hillard equation, also known as *Smoluchowski equation.*

For an infinitesimal fluctuation in composition we expand the free energy up to second order in fluctuation to obtain linearized diffusion equation of the following form,

$$\frac{\partial(\delta\phi(\mathbf{r}))}{\partial t} = D\nabla^2 \left\{ -K\nabla^2 + \left(\frac{\partial^2 f}{\partial \phi^2}\right)_{\phi=\phi_0} \right\} \delta\phi(\mathbf{r})$$

(4)

Since it is a linear differential equation and exactly solvable in Fourier space, one finds a time dependent solution for the order parameter as

$$\delta\phi(\mathbf{k},t) = \delta\phi(\mathbf{k},0) e^{-\omega(k)t},$$

(5)

where

$$\omega(k) = DKk^2 \left[ k^2 + K^{-1}\left(\frac{\partial^2 f}{\partial \phi^2}\right)_{\phi=\phi_0} \right].$$

As $\left(\frac{\partial^2 f}{\partial \phi^2}\right)_{\phi=\phi_0} < 0$ at very short time after the quench, $\omega(k)$ becomes negative for $k \to 0$. Hence an initial fluctuation grows in time for $k < k_c$ exponentially. In the forthcoming section we present a concise review of the various theories of spinodal decomposition.

## II. THEORY BEYOND CAHN AND HILLARD

Despite its elegance and simplicity, the analysis proposed initially by Cahn and Hilliard has many limitations. The most important limitation is its failure to describe the coarsening of the length scale of the domains with time. This has been partly rectified by a theory developed by Langer, Bar-on and Miller **[7, 8]** among others. The predicted long time decay is exponential as expected in any linear dynamical theory, but now the rate has more complex wave number dependence.

As noted by Kawasaki in his seminal paper **[9],** the situation of spinodal decomposition can be different for liquid binary mixtures compared to metallic alloys or other solid state. In binary mixtures, the rate of phase separation can thus be much higher giving rise to the situation where a transient growth can contribute significantly and thus lowering the importance of the long time dynamics. Using a generalized non-linear diffusion equation and ideas from mode coupling theory, Kawasaki has obtained the following elegant expressions for the decay rate $\omega(k,t)$ and the dynamic structure factor $S(k,t)$,

$$\omega(k,t) = A(\kappa^2 - k^2) t^{1/3} \text{ and}$$

$$S(k,t) = \exp(\frac{3A(\kappa^2 - k^2)}{4} t^{4/3}) \tag{6}$$

where A is a material constant consisting of viscosity and a few other parameters. According to Kawasaki's theory the growth may be faster than exponential at an early time.

Subsequently, in an independent study Suzuki **[10-15]** arrived at similar conclusions through a simpler approach. Suzuki has divided the entire progression of phase separation into three distinct regimes. First one is an exponential growth regime which can be understood easily as the broadening and separation of the probability distribution due to sliding down from an instability point or free energy maximum **[16, 17]**. Temporal evolution of fluctuation in initial regime is governed by the balance between drift term in linear part and diffusion term in the non linear part in the Fokker Planck equation. In the scaling or intermediate regime linear and nonlinear part in the drift term are responsible for the fluctuation. And finally diffusion term can take over in longer time scale. The description of sliding down from an instability point can lead to a divergence of second moment and cannot be connected to the final linear regime as described by Cahn and Hilliard. Hence Suzuki has proposed an intermediate regime and has termed it as the scaling regime where phase separation should be non-exponential.

### III.　OVERVIEW OF EXPERIMENTAL STUDIES

In several important articles, Goldburg **[18, 19]** and Knobler *et al.* **[20]** clearly demonstrated that phase separation of binary liquid mixture can occur at least through two distinct stages. At the beginning long wavelength delocalized concentration fluctuations grow exponentially until they form droplets which lie in the metastable states. Subsequently the droplets grow in size with a coarsening effect. They also found a prominent crossover from power law to linear domain growth indicating crossover from diffusive to viscous region **[18, 19].**

In another notable study, Goldburg *et al.* **[21]** reported a light scattering study of the dynamic behavior of a binary mixture of methanol and cyclohexane undergoing phase separation near the critical point. These intensity measurements in the early stages of the phase separation process found striking conformity with the linearized spinodal decomposition theory of Cahn. The lifetime of fluctuations in the spinodal region was also measured by Goldburg et al.

In a time resolved light scattering experimental study, Hashimoto *et al.* **[22, 23]** established an initial exponential increase in scattering intensity and subsequent deviation to power law regime for a binary mixture of Polystyrene (PS) and Poly vinyl methyl ether (PVME) [**22**]. This study had been performed far away from critical regime with a 70:30 mixture of the two polymers and well into the unstable region of phase diagram.

According to the studies described above, the linear theory is certainly capable of describing the early stage of SD. In the study of Hashimoto *et al.* **[22]** the initial exponential regime was found to continue only up to 20 to 30 minutes after initialization of phase separation for polymers of comparatively large size.

In this work we have used mass and radius of argon which are comparatively smaller with respect to that of polymers mentioned in the above experimental studies. Hence the time scale presented in our simulation study is quite small compared to the experimental study. In one of our earlier studies we have shown that time scale is weakly dependent on mass while controlled largely by the size of the polymers **[24]** as well as respective viscosities **[25]**. In the study of Hashimoto *et al.*, the coarsening effect (Oswald ripening process) has soon appeared resulting in increase of intensity with time departed from exponential behavior. This departure is found to be dependent on the depth of temperature quench.

In one more interesting study, Nishi *et al.* **[26]** have shown from the spin-lattice relaxation time T$_1$ in pulse NMR experiments that total reduction in signal $S_{NMR}(t)$ from polyvenyl methyl ether (PVME) in the polystyrene (PS) phase during spinodal decomposition has varied exponentially with time where $S_{NMR}$ is defined as,

$$S_{NMR}(t) = \left[ \int_{\frac{\pi}{2q_m}}^{\frac{3\pi}{q_m}} (c(x) - c_0) dx \right]^3 \sim \exp^{3R(q_m)t} \tag{7}$$

with c(x) denoting the composition at position x. The prediction by Nishi *et al.* is remarkably in good agreement with linear spinodal decomposition theory of Cahn-Hilliard **[1-5]**.

In case of large quench depth the exponential growth of time resolved scattering intensity has been found to initiate early. But, on a short time scale the linearized Cahn-Hillerd formulation is indeed valid **[1-6]**. In our present study, for quench in deep regions this crossover becomes weaker and the power law dominates over the initial exponential regime. This is the most important result of our numerical study and is in significant agreement with the above mentioned experimental studies.

## IV. OVERVIEW OF EARLY SIMULATION STUDIES

Because of the heavy computational requirement (large system size, long trajectory) phase separation kinetics is highly nontrivial to study through atomistic molecular dynamics simulation **[27-39]**. Most of the studies have employed a coarse grained approach, often directly treating the dynamics at the order parameter level. In contrast to the prediction of Kawasaki, several studies have suggested a linear growth of the domains **[29-34]**. The first ever simulation study of phase separation kinetics has probably done by Ma *et al.* **[29].** Later in a study, Laradji *et al*. also have found a small time linear growth of domain **[31].**

In the recent MD simulation study by Das *et al.* **[33]**, kinetics of phase separation in a binary LJ liquids has been demonstrated beautifully using a coarse-graining procedure. This study has confirmed the linear growth law in the viscous hydrodynamic regime. The morphological similarity of coarsening in solids and liquids has also been quantified.

In another simulation study **[37]**, we have demonstrated the inherent structures of structure breaker binary mixture to exhibit a wide range of structures ranging from nucleation to spinodal decomposition and the energy landscape view of non-ideality in structure breakers obtained from inherent structure analysis. The main finding is the interesting correlation between the energy of inherent structure and the nature of the microscopic mode of phase separation in the inherent structure.

The organization of the rest of the paper is as follows. We present an overview of theoretical (in Sec. II), experimental (in Sec. III) and simulation studies (in Sec. IV). In Sec. V, we discuss the simulation details. In Sec. VI, we present the detailed numerical results obtained from our simulations. Finally in Sec. VII we present the comparison of our system with real system and finally in Sec. VIII we briefly discuss results along with concluding remarks about future problems.

## V. MODEL AND SIMULATION DETAILS

A total number of 5324 Lennard-Jones particles is incorporated inside a cube of box length 18.433 with periodic boundary condition. We have checked by carrying out simulations with smaller system sizes so that the basic characteristics do not change much when system size is changed. The ratio of the composition of the two components was kept fixed at 60:40 which is away from critical point composition (1:1).

The phase behavior of a binary mixture is largely determined by the discrimination among the interaction parameters. One can write,

$$w_{AB} = \frac{\varepsilon_{AA} + \varepsilon_{BB} - 2\varepsilon_{AB}}{2} \tag{8}$$

where $\varepsilon_{AA}$, $\varepsilon_{BB}$ and $\varepsilon_{AB}$ describe interaction energy between two A, two B and one A with one B particles, respectively. All three interactions such as solute-solute, solvent-solvent and solute-solvent are expressed by the Lennard-Jones (12-6) potential,

$$U_{ij} = 4\varepsilon_{ij}\left[\left(\frac{\sigma_{ij}}{r_{ij}}\right)^{12} - \left(\frac{\sigma_{ij}}{r_{ij}}\right)^{6}\right]. \tag{9}$$

Here *i* and *j* denote any two particles. For simplicity, diameter (σ) and mass (m) for both solute and solvent atoms have been set to unity. The interaction strengths are $\varepsilon_{AA} = 1.0$, $\varepsilon_{BB} = 0.5$ and $\varepsilon_{AB} = 0.3$. In our present binary model, the Lennard-Jones interaction strength between A and B ($\varepsilon_{AB} = 0.3$) is less than both $\varepsilon_{AA}$ or $\varepsilon_{BB}$ i.e $w_{AB}$ is positive such that they favor phase separation and hence it corresponds to a "structure breaking" (SB) binary mixture **[37-39]**.

We have determined the critical temperature to be 2.3 for the chosen parameter values. At the specified composition (60:40 mixture), the coexistence temperature is at $T^* = 1.6 \pm 0.2$ as determined by varying the quench depth. Initially the system is equilibrated for 3ns at a high temperature $T^* = 5$.

In order to understand the kinetic behavior we have induced a shallow quench from $T^* = 5$ to $T^* = 0.9$ as well as a deeper quench to $T^* = 0.5$. We have quenched to lower enough temperature to avoid the solid gas coexistence line and also to ensure that we are in spinodal region far from nucleation regime **[40]**. Temperature is controlled by rescaling the velocities. We have also performed the same quenches but at different density values much lower than the coexistence line of LJ fluid solid transition to be sure that crystallization does not become an issue of concern. The equations of motion are integrated using Leap-frog algorithm with time step of 0.004 ps. *All the results presented here are averaged over 10 independent trajectories.*

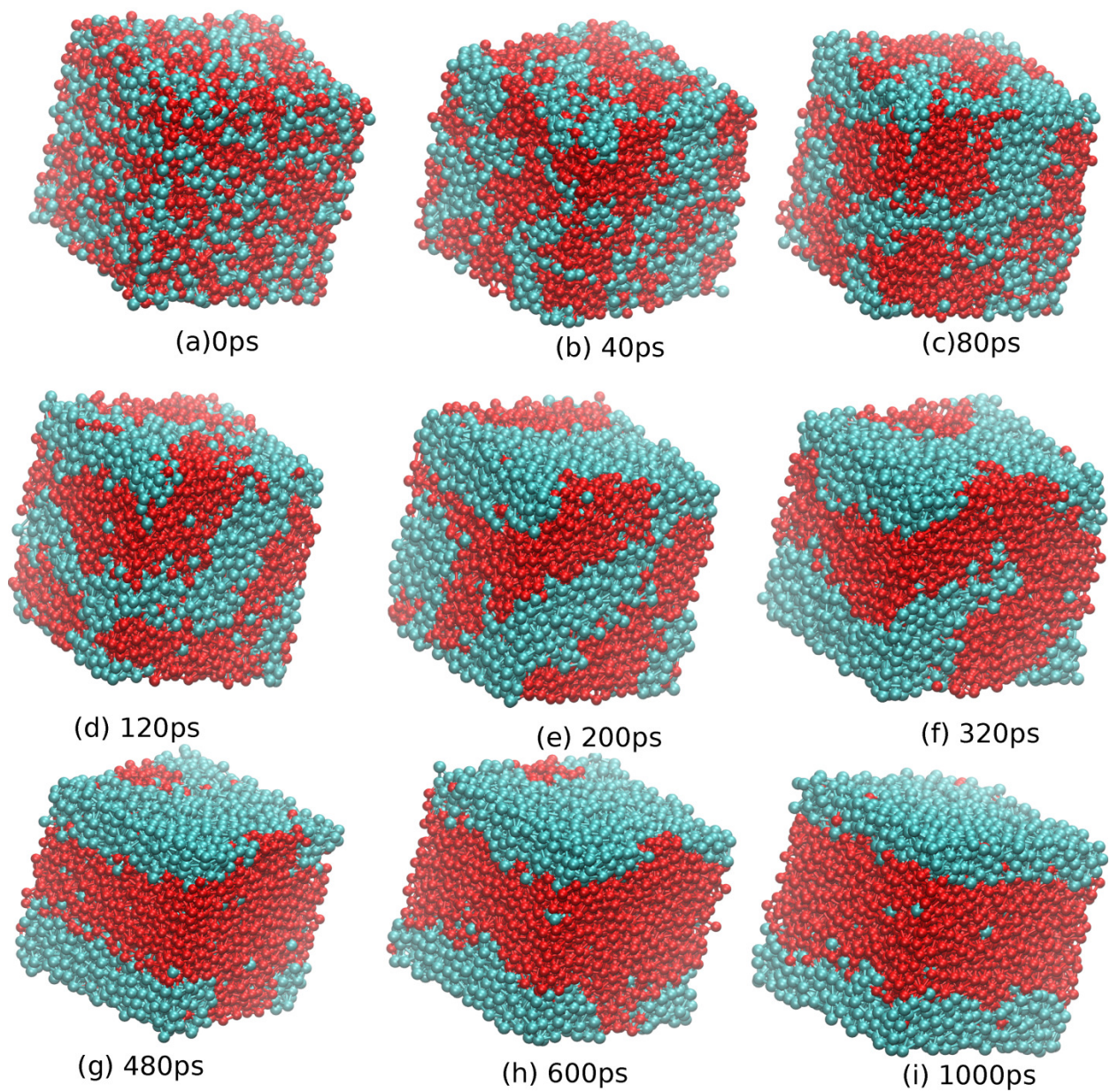

**FIG. 1. Representative snapshots from simulation showing time evolution of phase separation. The solvent (A) and solute (B) particles are represented by cyan and red spheres, respectively. (a) At initial time (t = 0), the system is seemed to be homogeneous. (b) At t=40ps small domains are getting formed. (c) At t = 80ps, phase separation proceeds rapidly with the increase in domain fronts as well as inter-connecting the smaller domains. (d) At t = 120ps, rapid process is almost completed and now the coarsening is taking over to tune the final form of domains. (e) At t = 200ps (f) t = 320ps (g) t =480ps (h) t = 600ps and (i) t = 1000ps, coarsening phenomenon is found to be in progress. A particles in B rich phases slowly move to a nearby A rich regions and vice versa.**

Fig. 1 shows several evolution snapshots obtained directly from simulation trajectories at different times. The solvent (A) and solute (B) particles are marked as cyan and red spheres, respectively. Snapshot at the initial time t=0 [as in **Fig. 1(a)**] corresponds to the homogeneous state immediately after the quench. The snapshots at 40 ps, 80 ps, 120 ps, 200 ps, 320 ps, 480 ps, 600 ps and 1000 ps [from **Fig. 1 (b)** to **(i)** respectively] show the spatial and temporal evolution of phase separation through spinodal decomposition.

## VI. RESULTS

### A. Spatial correlation and temporal dynamics of phase separation

We show the emerging spatial correlation in the system with respect to time in **Fig. 2(a)** and **Fig. 2(c)**. To define the correlation function we use resemblance of the binary mixture with that of Ising model. We define the correlation function such that it becomes unity when all the neighbors at a distance r around a central particle are of same type signifying phase separation. With increasing inter-particle distance the correlation function decays as different domains start appearing.

At initial times the correlation length is small as the islands of different particles are small and well separated. With increasing time, separate islands start growing and eventually spanning throughout the system. Average domain length is defined as the distance where the entire correlation function vanishes. The correlation function is defined as,

$$c_Q(r,t) = \frac{\langle Q(r,t) Q(r',t) \rangle}{\langle Q(r,t) Q(r,t) \rangle}. \tag{10}$$

As $Q(r,t)$ is defined for each particle, hence it can have only discrete values either +1 or -1. If the specified particle is of type A then we assign $Q(r,t)$ as 1 and if the particle under consideration is of B type then we assign $Q(r,t)$ as -1. **Fig. 2(a)** shows the spatial correlation for a quench from $T^* = 5$ to $T^* = 0.9$ where as **Fig. 2(c)** shows the same for a deeper quench from $T^* = 5$ to $T^* = 0.5$. The ripples in the correlation function in **Fig. 2(a) and Fig. 2(c)** are due to atomic size effect. As shown in **Fig. 2(a),** length scale of correlation increases with increase in time as A and B separate out from each other. We define the domain length, R(t) as the length scale at which this correlation vanishes.

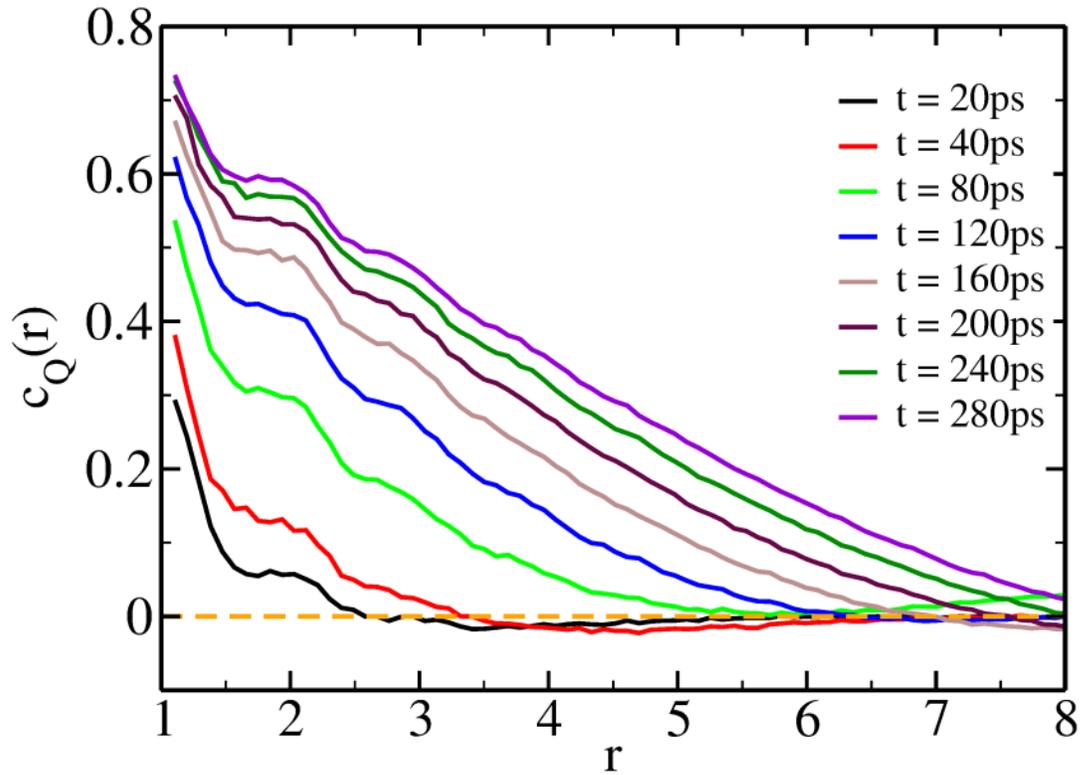

(a)

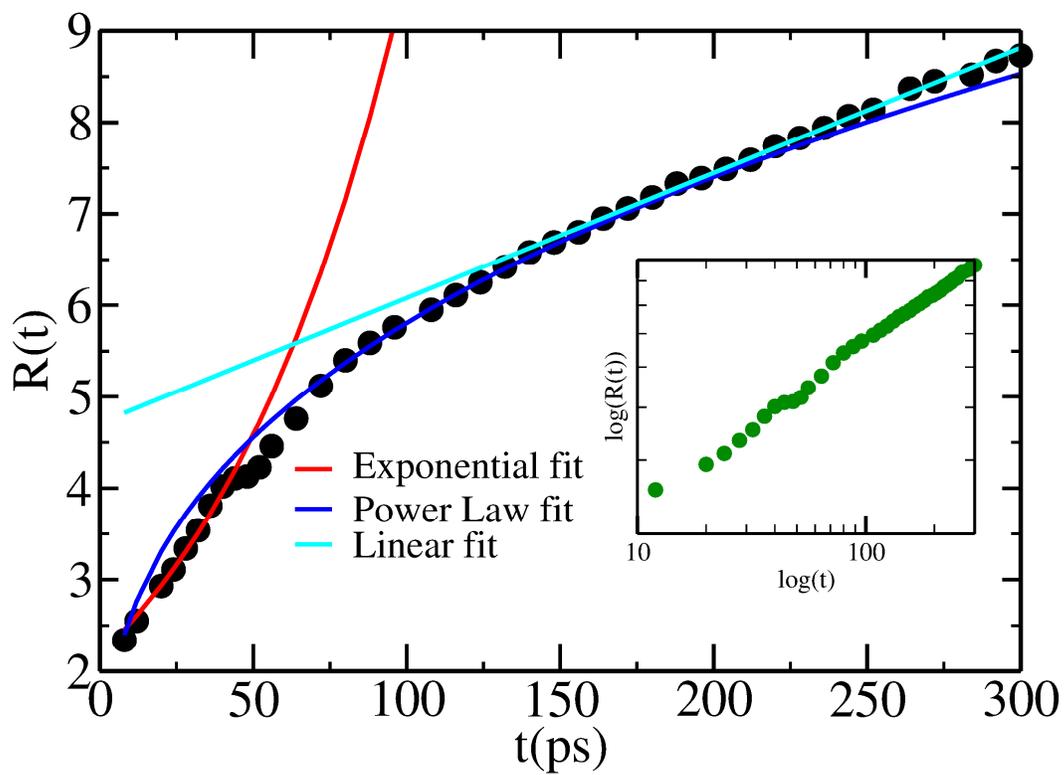

(b)

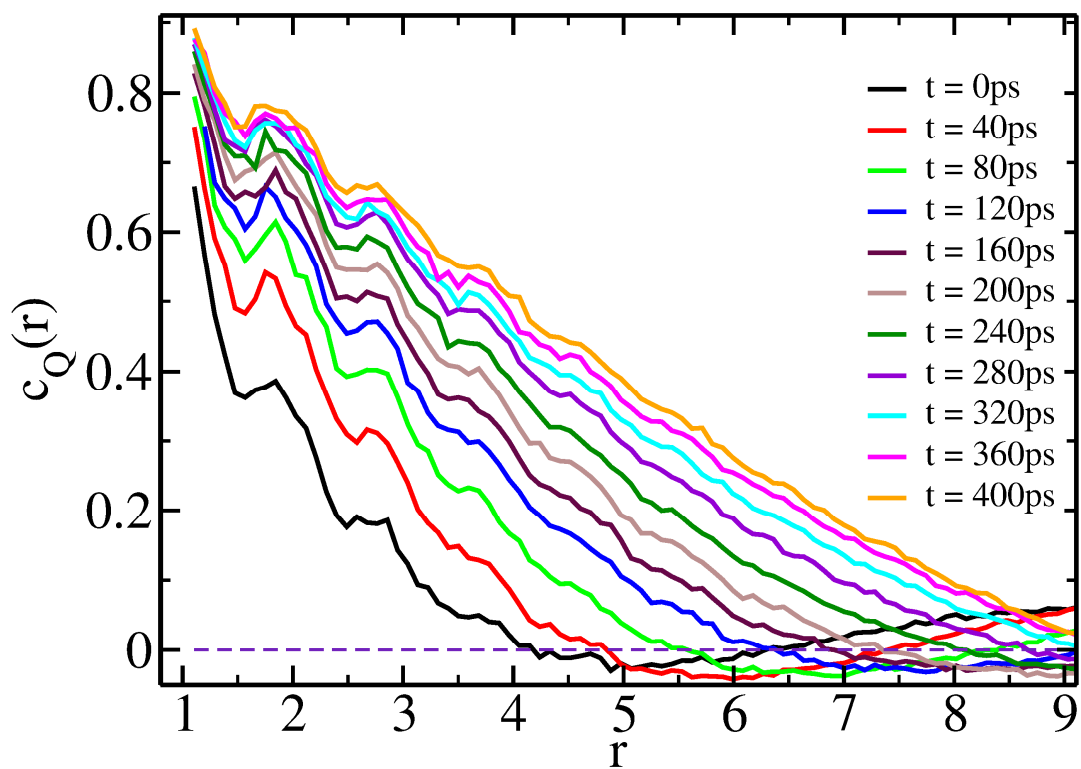

(c)

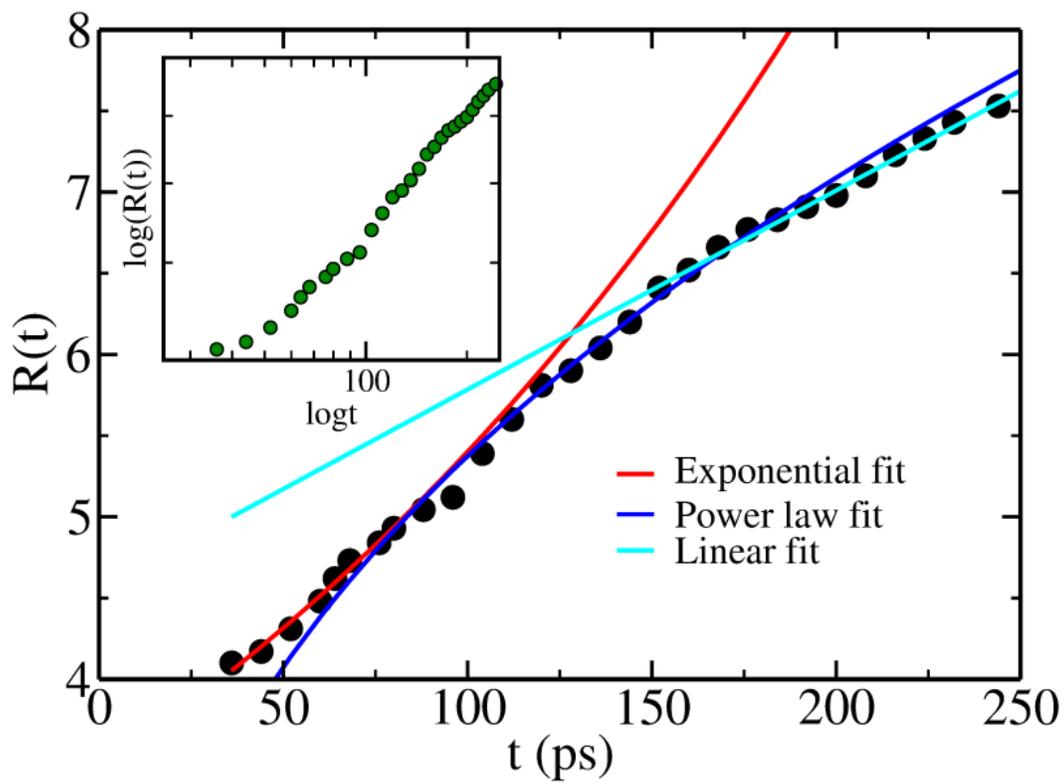

(d)

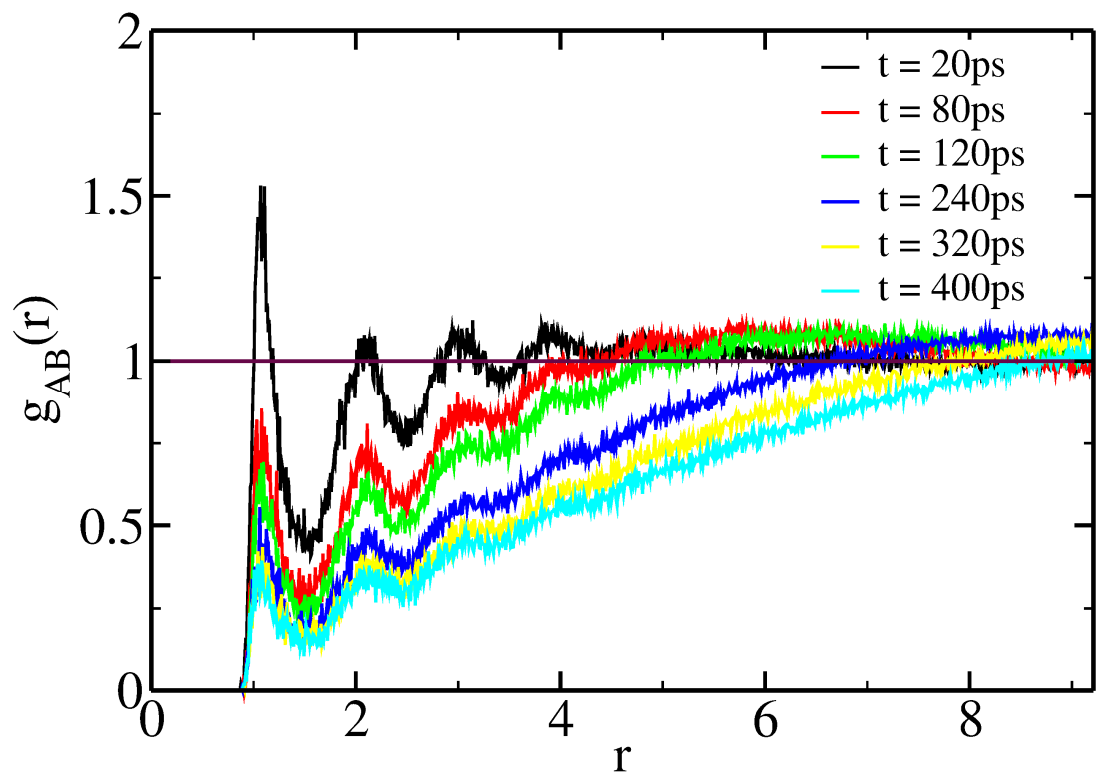

(e)

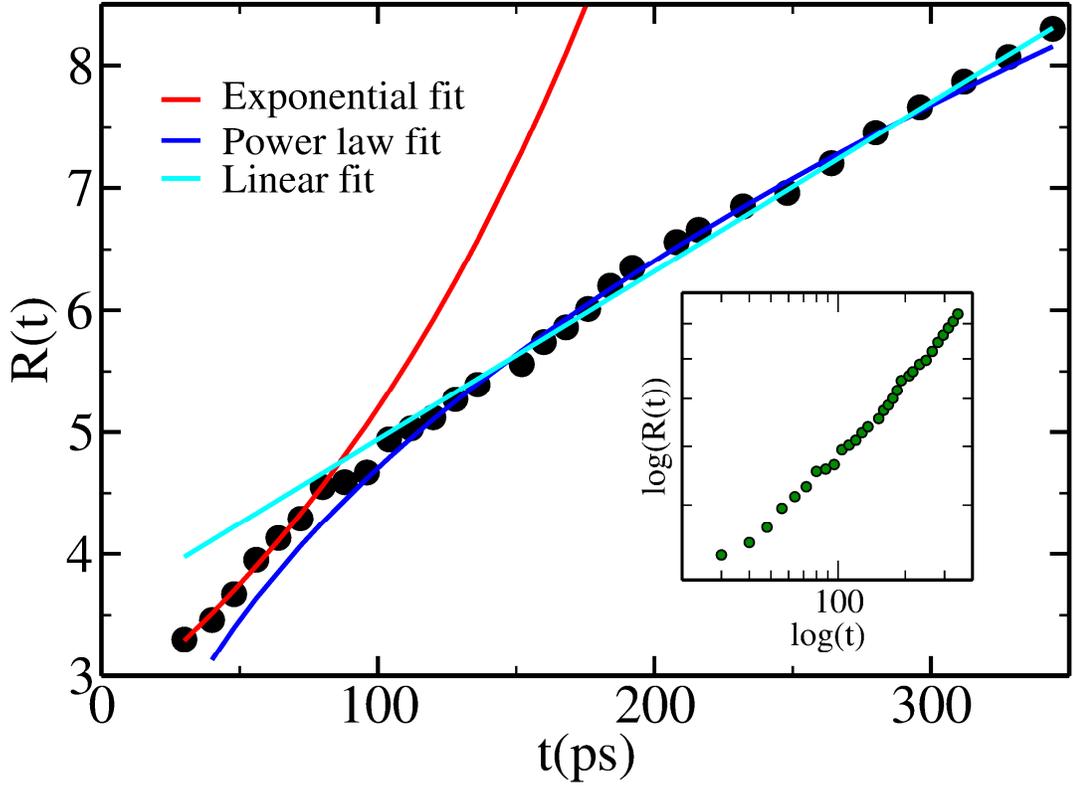

(f)

FIG. 2. Evolution of domain morphology with time for density 0.85. (a) Spatial correlation with time at quenched temperature $T^* = 0.9$. Ripples in the correlation function are due to molecular size effect. With increasing time A and B separate out from each other and length scale of correlation increases. Length scale at which this correlation vanishes is defined as domain length, R(t). (b) Corresponding domain length as a function of time. Initially exponential function shows a better fit than an overall power law at this temperature. Change over in the mechanism occurs around 50-100ps. Then a power law with exponent 0.36 shows a better fit. Inset shows the log-log plot. (c) Evolution of spatial correlation function with time at deeper quench temperature $T^* = 0.5$. (d) Corresponding domain length as a function of time. For this deeper quench, up to 30ps the process is slow enough and then it shows a considerable increase in domain length. The crossover to power law is gradual and occurs around 100ps. Inset shows the log-log plot. (e) Time evolution of $g_{AB}(r)$ against r at deeper quench temperature $T^* = 0.5$. Homogeneous bulk phase shifts to large length scale with time. (f) Length scale at which homogeneous bulk phase is reached is plotted against time. Similar three phase growth is evident. Log-Log plot shows the breakage between these regimes prominently. Both r and domain length (R(t)) are scaled to Argon diameter.

To get an insight into the underlying kinetics, we induce a shallow quench (to $T^* = 0.9$) and compare the results with a deeper quench (to $T^* = 0.5$) and investigate the comparative evolution of domain length. We look into the initial stage of phase separation in detail when the system slides down the hill along free energy surface starting from the unstable point. In **Fig. 2(b)** we show the evolution of average domain length with time for the quench $T^* = 5$ to $T^* = 0.9$ at density 0.85. Time evolution of domain length is showing three different regimes of growth explicitly. Initial exponential rise is evident up to 60ps in conformity with the theoretical prediction of Kawasaki, as well as Suzuki **[9-15].** Then there is a change in mechanism when there is a dropdown in the rate of growth. This can be seen by the kink in both R(t) vs t as well as logR(t) vs log(t). We fit initial 50ps to an exponential function of the form $y = a_0 \exp(-a_1 t)$ with time constant of 100ps. In the region (from 80 to 200ps) we find R(t) to follow a power law **[41]** of the form $y = a_0 + a_1 t^\alpha$ where the exponent $\alpha$ turns out to be 0.36 at quenched temperature $T^*$ *=0.9 which is in good agreement with the experimental finding of time resolved scattering intensity for PVME and PS liquid binary mixture reported by Hashimoto et al.* **[22]** while the time has to be scaled with respect to the much larger viscosity of the polymer binary mixture .

In **Fig. 2(d)** average domain length evolution has been shown for a quench from $T^* = 5$ to $T^* = 0.5$. For this deep quench cross over from exponential to power law is not sharp as in the case of $T^* = 0.9$. Initially up to 75ps we find a fast exponential increase in domain length with time scale of 83ps when fitted to an exponential function of the form $y = a_0 \exp(-a_1 t)$. Region from 100ps to 160ps is fitted with power law of the form $y = a_0 + a_1 t^\alpha$ and the exponent $\alpha$ turns out to be 0.42. Due to such deep quench exponential increase in domain length becomes too hard to capture as the power law sets in early.

As evident from **Fig. 2(b)** and **2(d)**, domain length steadily increases up to 150-200ps; subsequently any increment of domain length is slow and dominated by coarsening. The shift in the average domain length to large length scale at later time is a signature of coarsening and long range phase separation in the system.

**Fig. 2(e)** shows the radial distribution of B particles around A at different times during phase separation process at quenched temperature $T^* = 0.5$. With increasing time homogeneous bulk phase is achieved at increasingly large length scale and eventually inhomogenity spreads throughout the whole system. We plot length scale at which homogeneous bulk phase is obtained with time in **Fig. 2(f)**. It is interesting to note here that multistage dynamics is evident and is in well agreement with the analysis from spatial correlation as given by **Fig. 2(d)**.

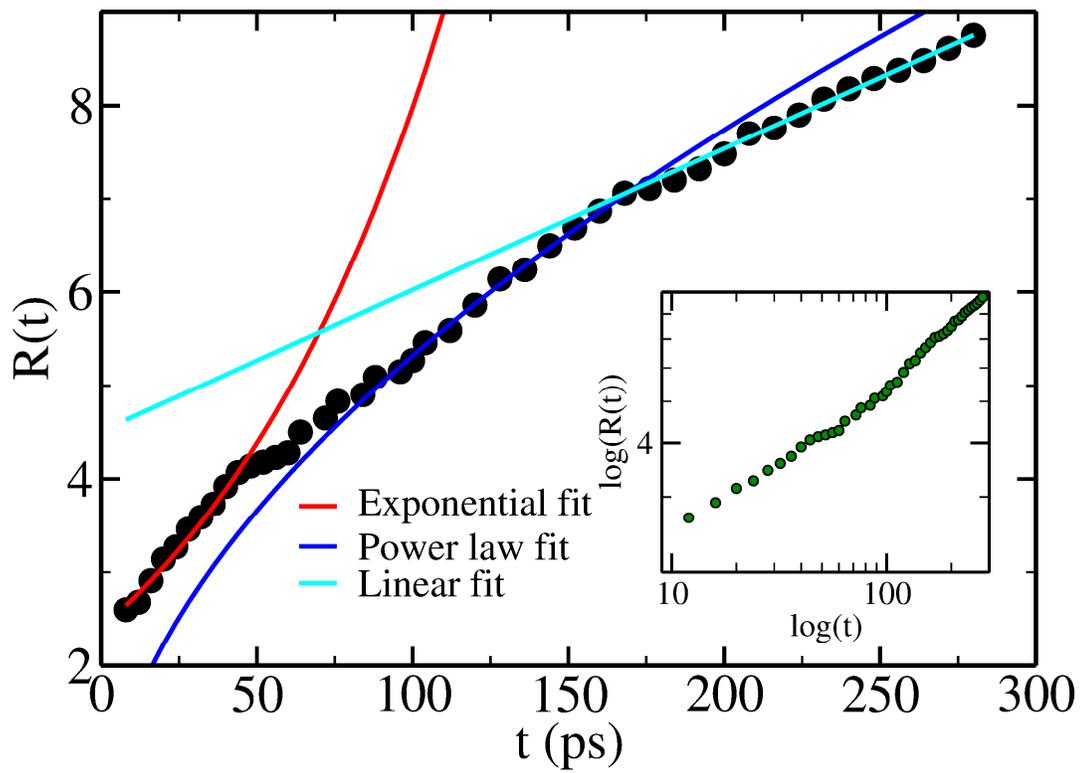

(a)

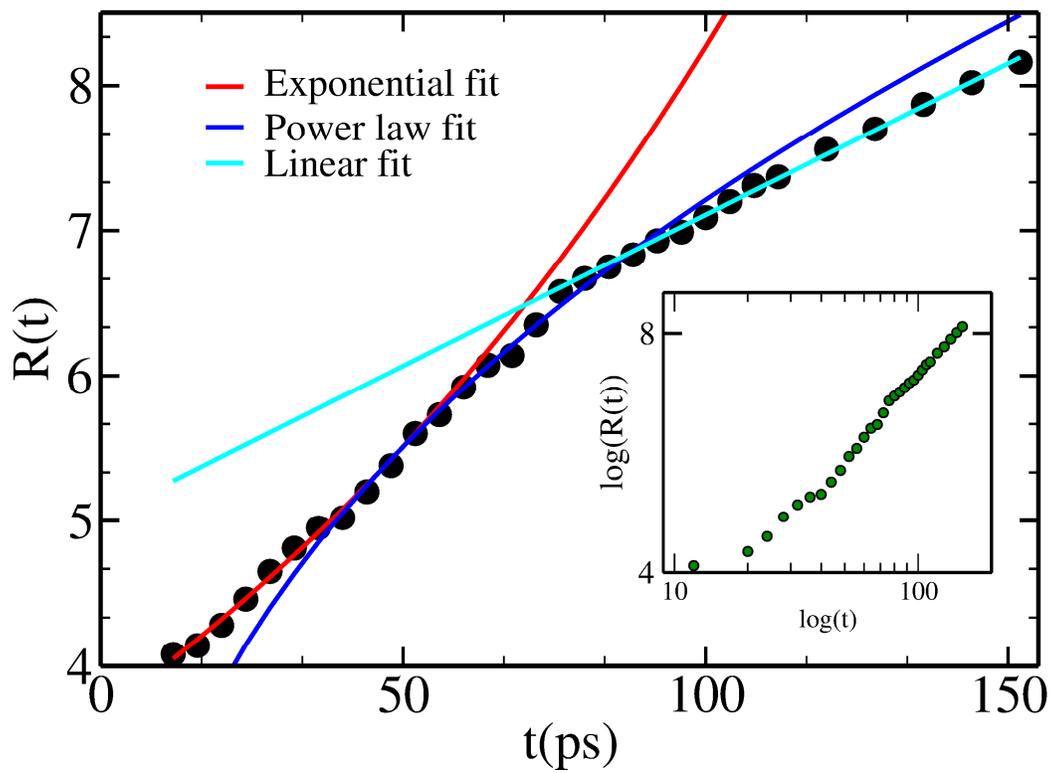

(b)

**FIG. 3. Characteristic domain length growth for density 0.75. (a) Plot shows the evolution of domain length at T$^*$=0.9. Early time of spinodal decomposition is fitted to an exponential function up to 50ps with a time constant of 83ps. Then Power law sets in. Up to 150ps the power law fits well with exponent 0.50. Inset shows log-log plot. (b) Evolution of domain at T$^*$= 0.5. Initial exponential behavior is seen up to 50ps with time constant of 120ps. Inset shows the corresponding log-log plot which clearly signifies a multi-stage dynamics. At lower density due to higher diffusion of the particles, diffusive growth (power law) stays for a considerable time span.**

In order to verify the generality of this multi stage kinetics as well as the effect of crystallization we calculate the same spatial correlation at lower density 0.75. We confirm that crystallization does not affect our system at all during the kinetics of phase separation. **Fig. 3(a)** and **Fig. 3(b)** show evolution of average domain length for the quenches $T^* = 5$ to $T^* = 0.9$ and $T^* = 5$ to $T^* = 0.5$ respectively at density 0.75. Early time of spinodal decomposition for the quench from $T^* = 5$ to $T^* = 0.9$ is fitted up to 50ps to exponential function with a time constant of 83ps. After that the power law fits well up to 150ps with exponent 0.50. In case of the quench from $T^* = 5$ to $T^* = 0.5,$ initial exponential behavior is found up to 50ps with a time constant of 120ps. The same trend of dependence over quench depth is present here also i.e., for deeper quench exponential to power law transition is mild to that of a shallow quench.

Dynamics of these regimes exhibit different quench depth dependence. Initial regime is sensitive to temperature gradient. If we increase the depth of temperature quench, the difference between this region and the power law region becomes smaller and eventually the power law dominates with deeper quench.

Time resolved light scattering studies by Hashimoto *et al.* **[22]** also reveal the same dynamical features. The power law regime has seemingly increased life span for a deeper quench depth as well as for lower density. The crossover from power law to linear coarsening regime becomes gradual afterwards. Power law regime is also governed by diffusivity of the medium. In a system where density is quite low diffusion dominates and we find an increased exponent 0.5 in the power law stage of dynamics. In order to get an idea of the time scales involved in microscopic flitting of A-B contact dynamics, we monitor the probability of continuous time along which an A-B contact remains. It is quite interesting to note that during simulation we observe the time scale involved in microscopic flitting of A-B contact dynamics to be around 10-20ps only and this time scale has no temperature dependence as such.

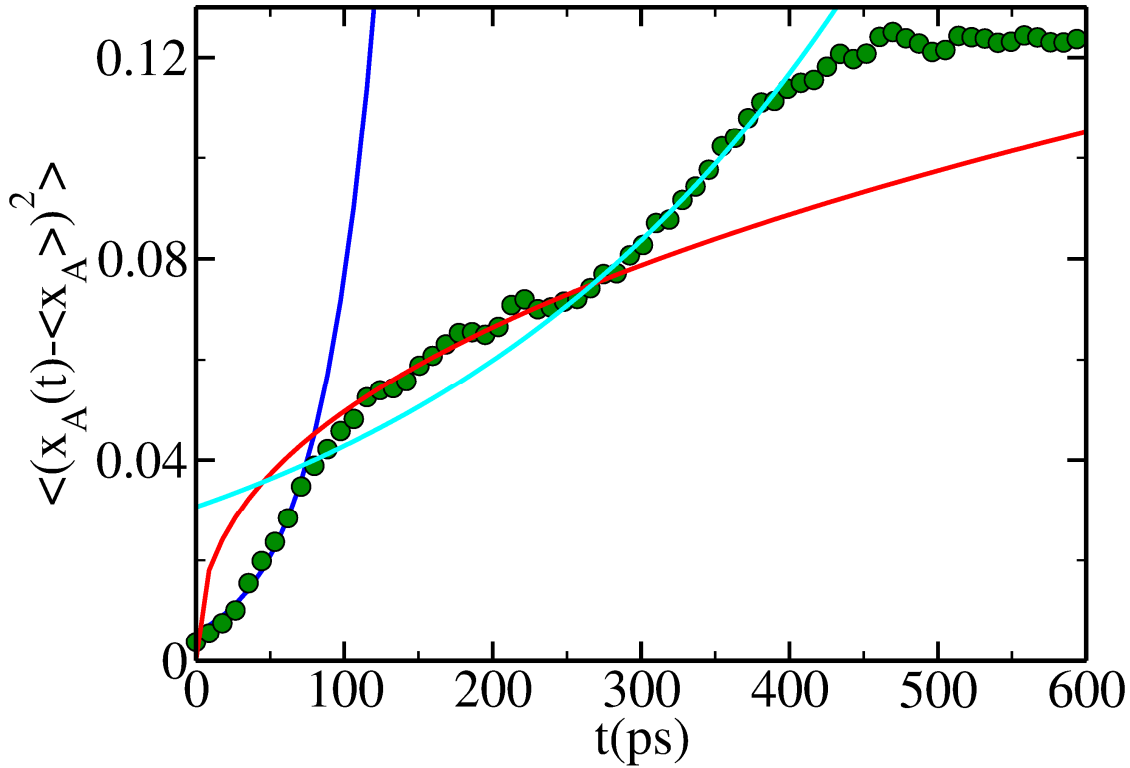

**FIG. 4. Local fluctuation in composition of A averaged over all the grids at quenched temperature $T^*=0.9$. Phase separation is almost over by 500ps. Three regimes of phase separation is also an eminent feature here.**

**Upto 100ps kinetics follows a fast exponential growth. After that a slow crossover region in the time scale of 100-200 ps follows and finally a linear law takes over to the completion of phase separation.**

This multi scale dynamics of phase separation is fairly evident when we address the problem from a different point of view. We coarse grain the system of our interest into small grids and evaluate locally the second moment of composition variation of solvent A and averaging over all the grids. **Fig. 4** shows the same for the quench $T^* = 5$ to $T^* = 0.9$. Upto 100ps there is an initial exponential regime with time scale of 50ps followed by a break and a slow increase in fluctuation upto 200ps. This stage when fitted to a power law of the form $y = a_0 + a_1 t^\alpha$ gives a value of the exponent $\alpha = 0.40$. Finally, there is change over to coarsening regime. In final stage of growth, the increase in fluctuation shows a high rate. This stage indicates the reorganization of different domains as they get connected to each other eventually, which is also evident from snapshots at later stages of growth (**Fig.1**).

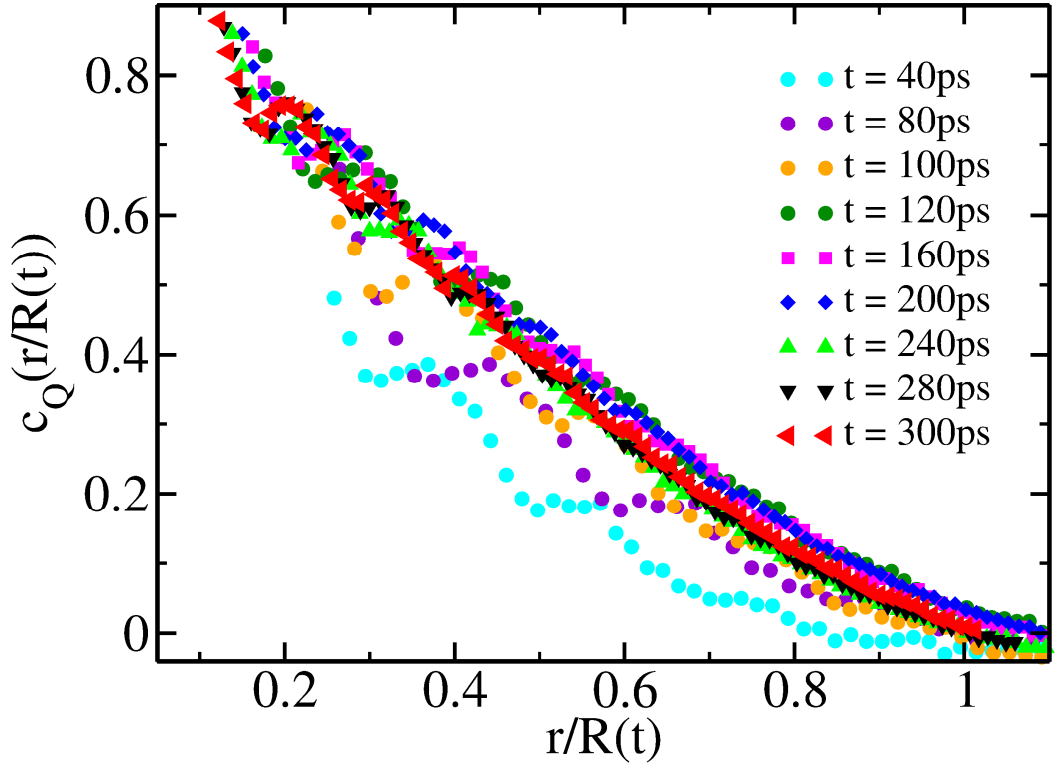

**FIG. 5.** Spatial correlation function plotted with r (scaled with domain length) for quenched temperature $T^*=0.9$. All the spatial correlation functions follow a master plot from time scale 100ps onwards. Interestingly spatial correlation in the time range 0-100ps show significant deviation from the master plot. The deviation signifies that the onset of the coarsening phenomena appear only after 100ps time scale.

**Fig.5** illustrates certain interesting features about the spatial correlation function with respect to r for the quenched temperature $T^* = 0.9$. The distance between two particles r is scaled by the average domain length R(t) of the corresponding time t. This figure shows that the spatial correlation follows a single master curve which is a signature of typical coarsening phenomenon depicting the common underlying dynamics. Interestingly spatial correlation for the plots at 40ps, 80ps and 100ps show considerable deviation from the master curve and all the spatial correlation functions from 100 ps onwards follow the same master plot. The disappearance of the deviation signifies the onset of the coarsening phenomena that appears only after 100ps time scale. Coarsening has often been discussed in literature in the context of phase separation **[36]**.

### B. Heterogeneous local dynamics and initial composition dependence

**Fig. 6** depicts the correlation of order parameter in different regions of the system with respect to time. The correlation function is defined as

$$c_{\phi\phi}(r,t) = \frac{\langle \phi(r,t_0)\phi(r,t_0+t) \rangle}{\langle \phi(r,t_0)\phi(r,t_0) \rangle} \quad (11)$$

Information about heterogeneous phase separation kinetics is revealed by coarse graining the cubic box into 27 small grids and evaluating local order parameter in each grid with time. Regions initially enriched in A or B ($\phi \geq 0.40$ or $\phi \leq -0.35$) grow very fast almost within 300 - 400ps with very less fluctuation in order parameter with time. The initial regime up to 100ps fits exactly to an exponential function and subsequently deviates from exponential behavior. For a quench depth of $T^* = 5$ to $T^* = 0.5$ time constant becomes 116ps for B rich region and 153ps for A reach region. But regions with almost equal mole fraction of A and B ($\phi \approx 0$) grow relatively slow. Some of these regions serve as channels for the transport of collective particles during 2$^{nd}$ stage of growth and ultimately form domain boundaries where fluctuation is long range in time. These fluctuations are due to creation of surface between two phases with unfavorable interaction and also due to collective transport of certain type of particles. Regions with almost equal distribution of A and B show a slow creation of concentration gradient which is the trademark of "uphill diffusion". Kinetics shows an initial exponential behavior. All the localized grids show this exponential behavior in general. Interestingly, there is a regime around 100ps in A rich or B rich regions where we find two intersections to the fitted exponential curve, one for a changeover from fast exponential to a slow power law dynamics due to which correlation drops rather slowly and another in the regime of 300-400ps where kinetics again becomes fast enough to cross the fitted exponential after the crossover due to reorganization of the domains created during the initial stages. This feature seems to be evident in all the time dependent properties we are interested in.

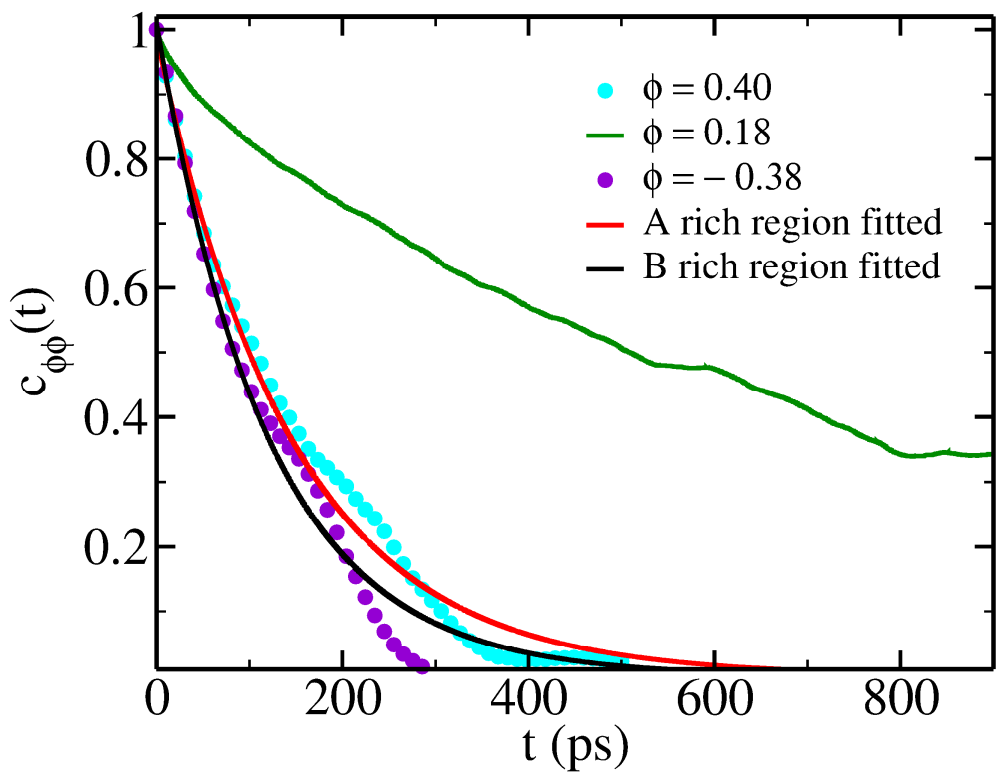

(a)

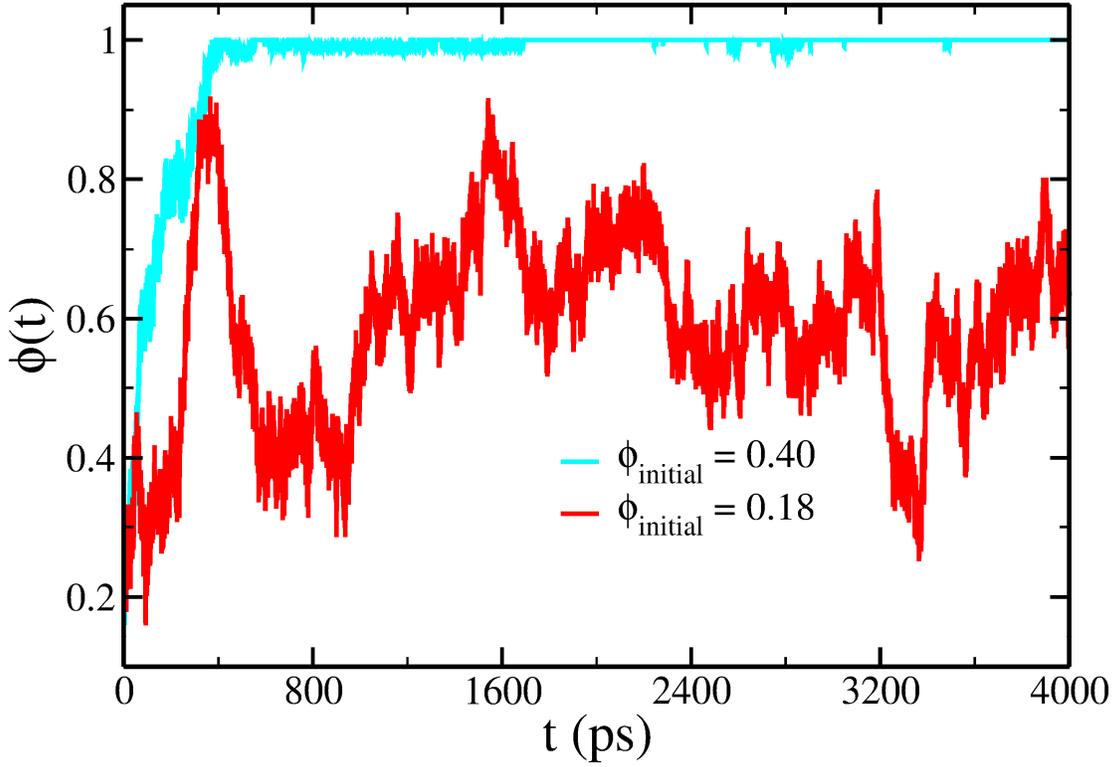

(b)

FIG.6. Heterogeneous phase separation kinetics captured through evolution of local correlation of order parameter with time at $T^* = 0.5$ and density 0.85. (a) Plots of temporal correlation function for the regions having initial order parameters 0.40, 0.18, -0.38. Regions initially enriched in A or B grow comparatively faster (almost within 300-400ps) with less fluctuation in order parameter whereas the regions having almost equitable mole fraction of A and B seem to grow relatively slow. (b) Plots of representative order parameters 0.40, 0.18 for the corresponding regions with respect to time. The fast growing regions are rich in either A or B phase.

### C. Temporal evolution of clusters of A and B particles

We have also analyzed the segregation of phases from the dynamics of cluster formation. We follow the traditional way to define a cluster of non bonded particles. As two types of particles eventually get separated, we can define cluster comprising of each kind of particles. We define the cutoff distance to find all the nearest neighbors as the first minimum of radial distribution function. We define a particle to be a member of a particular type of cluster (Either A or B), if that particle has 10 nearest neighbors of its own kind. **Fig. 7(a)** shows the decay of the number of clusters consisting of A particles with time for a quench from $T^* = 5$ to $T^* = 0.9$ and density 0.85. Initial region up to 100ps is fitted to exponential with a time constant of 83ps. Inset shows the power law stage with an exponent of 0.37. Exponential form breaks down at 80ps timescale onwards.

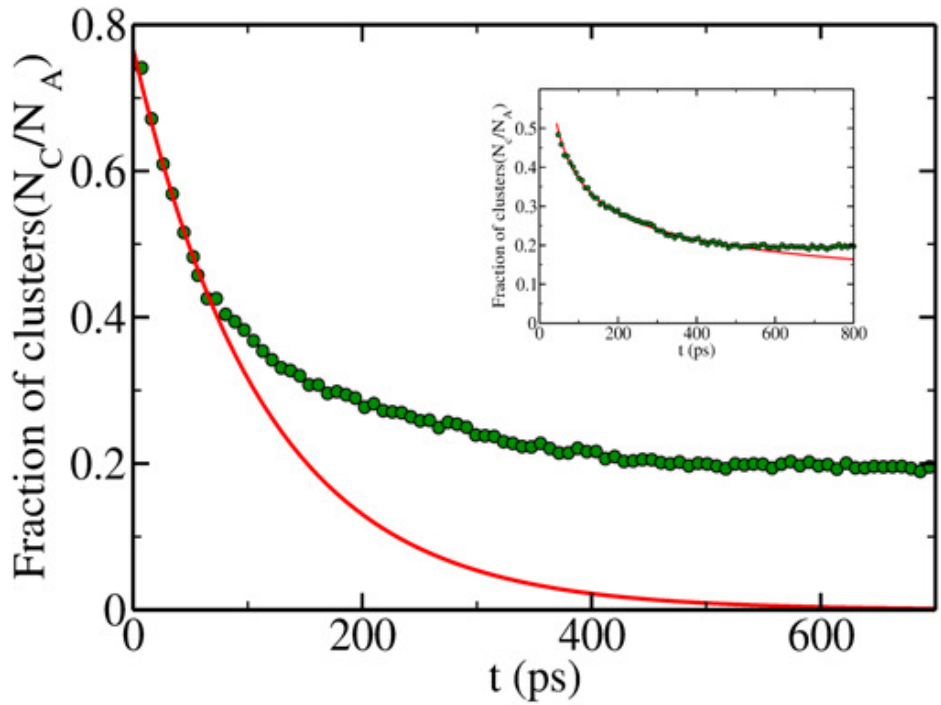

(a)

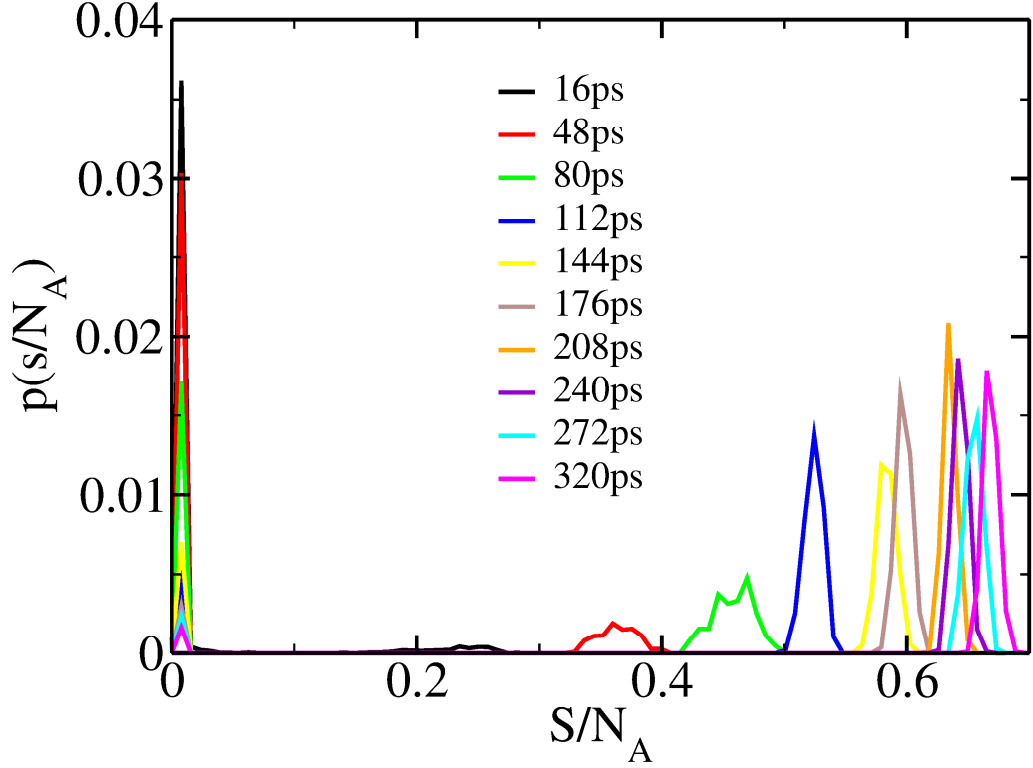

**(b)**

FIG. 7. Time dynamics of fraction of clusters for the quench from $T^*=5$ to $T^*=0.9$ at density 0.85. (a) Number of clusters of A type of particles also shows an initial exponential decay with time constant of 83 ps. Inset shows the power law decay after 100ps. The exponent turns out to be 0.37. (b) Cluster size (S) distribution of A type of particles at different times. Initially probability distribution is concentrated on single particle and eventually up to 320ps significant growth of cluster is found. Shifting of probability distribution to larger S values is the signature of the emergence of domain. Single particle probability decreases with increasing time and distribution shifts towards large cluster size signifying the segregation of phases.

**Fig. 7(b)** shows distribution of cluster size (S) at different times for A type of particles. Cluster size is scaled to the largest possible cluster size of A particles. Shift of probability distribution to larger S values signifies emergence of a domain. Significant growth of cluster is visible up to 320ps. As usual initial probability of cluster size manifests itself as single particle distribution. With increasing time, single particle probability density diminishes and distribution shifts towards large cluster size pointing segregation of phases.

## VII. COMPARISON WITH REAL SYSTEMS

As mentioned several times earlier, present model calculations employ parameters that induce phase separation when quench below co-existence temperature. Specific nature of the constituents and use of parameters for argon are certainly not useful for comparison with experiments. In particular, one would like to have predictions regarding the time scales that are expected to represent different kinetic regimes. As for example, one would like to know the time scale for crossover from exponential to power law & subsequent crossover from the Suzuki's prediction to that of Kawasaki.

We have therefore adopted the usual procedure to convert the simulated time to real time by scaling with the ratio of viscosity for the real and model system **[25]**. This ratio is found to be 1000 for water-Lutidine binary mixture. We therefore predict for shallow quench the interesting dynamical range to be from 50 ns to 500 ns with the crossover region to be expected from 100 ns to 200 ns range. The intermediate to long time dynamics may be accessed optically by time domain experiments and laser pulses.

## VIII. DISCUSSION AND CONCLUDING REMARKS

Kawasaki-Suzuki scaling theory predicts that the initial fast exponential-like growth embodied in the Cahn-Hilliard theory is to be replaced by a much slower, non-exponential long time growth. This regime is widely known as the scaling regime. In fact, for phase separating liquid binary mixtures, one may expect three different regimes, although any quantitative analysis of the relative weights of all three does not seem to exist, and also the crossovers are not well studied. In liquid binary mixtures, the kinetics is much faster than that in binary alloys, and the time scales are found to be shorter. The three stages should be observable if a temperature quench can be accomplished by some fast means, like laser excitation.

We now briefly summarize the main results of this work. We have carried out atomistic molecular dynamics (MD) simulation of a structure breaking binary liquid mixture, by varying the depth of quench and also the total density of the system. The parameters are so chosen that the system is away from the critical point.

We have combined our MD simulations with a coarse grained multi scale modeling (CGMSM) where we divide the system into cubic boxes of smaller size and study the evolution of composition in each box. This CGMSM modeling allows us to capture both the length and the time scales of phase separation kinetics within the simulation box.

In the present study the homogeneous liquid system is quenched from a higher temperature ($T^*=5$) to two lower temperatures ($T^*=0.5$ and $T^*=0.9$) well below the coexistence temperature $T^* = (1.6 \pm 0.2)$ of the phase diagram.

We observe rich multi scale phase separation dynamics, summarized below.

(i) For the system parameters adopted, the initial growth is exponential up to 80-100ps. This region exhibits strong dependence on quench depth. Subsequently, a crossover regime appears rather sharply where the dynamics slows down considerably leading to coarsening through power law time dependence. We have observed three dynamic domains and two crossovers. As discussed, for realistic systems, this time can be scaled up by using the respective viscosity.

(ii) For deeper quench, the power law growth dominates over the initial exponential growth whose importance decreases markedly and the crossover from exponential to power law occurs at shorter times. We find that for the present parameter values of the liquid mixture, the initial rapid growth of structure formation is practically over within 200 ps which is followed by slow structural coarsening.

(iii) As mentioned repeatedly above, the present study uses argon-type model parameters which need to be aligned with real world systems. This can be done by scaling the relaxation time by the respective viscosity. When scaled by the respective viscosities, this initial time (50-200 ps) translates to 50-200 ns for water-lutidine binary mixture. Thus, our predictions can be tested by experiments. The slow growth may continue up to ms range.

(iv) Importantly, our course grain studies reveal that the phase separation kinetics is heterogeneous in the sense that the kinetics, even the nature, of initial phase separation, are found to exhibit different microscopic timescales in different regions of the system depending on the initial composition of the probed region, as further discussed below.

(v) The regions enriched significantly in one of the components exhibit a faster timescale of reorganization than in those regions where both components are present in nearly equal measure and the phase separation is nucleation mediated, leading monotonously to a completely phase separated equilibrium state. The dynamics of phase separation is slowest in regions that have equitable distribution of the two species, and one can observe the signatures of "up-hill diffusion" that is a trade mark of spinodal decomposition.

(vi) Phase separation dynamics is found to slow down considerably when the final quenched temperature is moved closer to the estimated critical point.

Observations of the present study could be probed in experiments.

## ACKNOWLEDGEMENT

We thank Dr. Alexei Goun (Princeton University) for many stimulating discussions. We acknowledge useful discussions with Professor S. K. Das and Professor S. Puri. BB thanks Department of Science and Technology (DST, India) and Sir J. C. Bose Fellowship for providing partial financial support.